\documentstyle[psfig]{article}   

\newcommand{\hide}[1]{}
\newcommand{\sequence}[1]{\left\langle#1\right\rangle}

\newtheorem{theorem}{Theorem}
\newtheorem{lemma}{Lemma}
\newtheorem{prop}{Proposition}

\begin{document}

\title{\Large Pairing Heaps with Costless Meld}

\author{Amr Elmasry \thanks{Supported by an Alexander von Humboldt Fellowship.}\\
Max-Planck Institut f\"{u}r Informatik \\
Saarbr\"{u}cken, Germany \\
\small elmasry@mpi-inf.mpg.de}
\date{}

\maketitle

\begin{abstract} \small\baselineskip=9pt
Improving the structure and analysis in \cite{elm0}, we give a variation of the pairing heaps that has amortized zero cost per meld (compared to an $O(\log \log{n})$ in \cite{elm0}) and the same amortized bounds for all other operations. 
More precisely, the new pairing heap requires: no cost per meld, $O(1)$ per find-min and insert, $O(\log{n})$ per delete-min, and $O(\log\log{n})$ per decrease-key. These bounds are the best known for any self-adjusting heap, and match the lower bound proven by Fredman for a family of such heaps. Moreover, our structure is even simpler than that in \cite{elm0}.
 
\end{abstract}

\section{Introduction}

The pairing heap \cite{fsst} is a self-adjusting heap that is implemented as a single heap-ordered multi-way tree. The basic operation on a pairing heap is the linking
operation in which two trees are combined by linking the root with
the larger key value to the other as its leftmost child. 
The following operations are defined for the standard implementation of the pairing heaps:

\begin{itemize}
\item{\it find-min.} Return the value at the root of the heap.
\item {\it insert.} Create a single-node tree and link it with the tree of the heap.
\item {\it decrease-key.} Decrease the value of the corresponding node. If this node is not the root, cut its subtree and link the two resulting trees.
\item{\it meld.} Link the two trees representing the two heaps.
\item {\it delete-min.} Remove the root of the heap and return its value. The resulting trees are then combined to form a single tree. For the standard two-pass variant, the linkings are performed in two passes. In the first pass, called the pairing pass, the trees are linked in pairs from left to right (pairing these trees from right to left achieves the same amortized bounds). In the second pass, called the right-to-left incremental-linking pass, the resulting trees are linked in order from right to left, where each tree is linked with the tree resulting from the linkings of the trees to its right.
Other variants with different {\it delete-min} implementation were given in \cite{e, f1, fsst}.
\end{itemize}

The original analysis of the pairing heaps \cite{fsst} showed an $O(\log{n})$ amortized cost for all operations.
Another self-adjusting heap that requires $O(\log{n})$ amortized cost per operation \cite{st2} is the skew heap. 
Theoretical results concerning the pairing heaps were later obtained through the years.
Stasko and Vitter \cite{sv} suggested a variant that achieves $O(1)$ amortized cost per {\it insert}. 
The bounds for the standard implementation were later improved by Iacono \cite{ia} to: $O(1)$ per {\it inset}, and zero cost per {\it meld}. 
Fredman \cite{f} showed that $\Omega(\log \log{n})$ amortized comparisons, in the decision-tree model, would be necessary per {\it decrease-key} operation for a family of heaps that generalizes the pairing heaps. Pettie \cite{p} proved amortized costs of: $O(\log{n})$ per {\it delete-min}, and $O(2^{2 \sqrt{\log \log n}})$ for other operations. Recently, Elmasry \cite{elm0} introduced a variant that achieves the following amortized bounds: $O(1)$ per {\it insert}, $O(\log{n})$ per {\it delete-min}, and $O(\log \log{n})$ per {\it decrease-key} and {\it meld}. See Table \ref{t1}.

\begin{table}[tb!]
\label{t1}
\caption{Previous results for upper bounds on pairing-heap's operations}
\begin{small}
\begin{center}
\begin{tabular}{|c|c|c|c|c|}
\hline
 & {\it insert} & {\it delete-min} & {\it decrease-key} & {\it meld} \\
\hline
Fredman et al. \cite{fsst} & $O(\log{n})$ & $O(\log{n})$ & $O(\log{n})$ & $O(\log{n})$ \\
\hline
Stasko and Vitter \cite{sv} & $O(1)$ & $O(\log{n})$ & $O(\log{n})$ & $O(\log{n})$ \\
\hline
Iacono \cite{ia} &  $O(1)$ & $O(\log{n})$ & $O(\log{n})$ & zero \\
\hline
Pettie \cite{p} & $O(2^{2 \sqrt{\log \log n}})$ & $O(\log{n})$ & $O(2^{2 \sqrt{\log \log n}})$ & $O(2^{2 \sqrt{\log \log n}})$ \\
\hline
Elmasry \cite{elm0} & $O(1)$ & $O(\log{n})$ & $O(\log \log {n})$ & $O(\log \log{n})$ \\
\hline
This paper & $O(1)$ & $O(\log{n})$ & $O(\log \log {n})$ & zero \\
\hline
\end{tabular}
\end{center}
\end{small}
\end{table}

Several experiments were conducted on the pairing heaps, either comparing its performance with other priority queues \cite{j,ms} or with some of its variants \cite{e,f1,sv}. Such experiments illustrate that the pairing heaps are practically efficient and superior to other heaps, including the Fibonacci heaps \cite{ft}. 

In this paper, we give a variation of the pairing heaps that achieves the best known bounds for any self-adjusting heap for all operations. Namely, our amortized bounds are: zero cost per {\it meld}, $O(1)$ per {\it find-min} and {\it insert}, $O(\log{n})$ per {\it delete-min}, and $O(\log\log{n})$ per {\it decrease-key}.
We describe the data structure in Section $2$, prove the time bounds in Section $3$, give possible variations in Section $4$, and conclude the paper with some remarks.

\section{The data structure}

Similar to the standard implementation of the pairing heaps, we implement our variation as a single heap-ordered multi-way tree. 
Since we perform the {\it decrease-key} operations lazily, a pointer to the minimum element is maintained.\\

The detailed implementations for various heap operations are as follows:

\begin{itemize}

\item{\it find-min.}
Return the value of the node pointed to by the minimum pointer.

\item {\it insert.}
Create a single-node tree and link it with the main tree. Update the minimum pointer to point to this node if it is the new minimum.

\item{\it decrease-key.}
Decrease the value of the corresponding node $x$. Update the minimum pointer to point to $x$ if it is the new minimum. Add $x$ to the list of decreased nodes if it is not a root.

\end{itemize}

We use the following procedure in implementing the upcoming operations:

\begin{description}

\item {$\; \; \; \;$ -\it clean-up:}

\begin{enumerate}

\item[i.] Perform the following for every node $x$ in the list of decreased nodes: Cut $x$'s subtree and the subtree of the leftmost child of $x$. Glue the subtree of the leftmost child of $x$ in place of $x$'s subtree, and add the rest of  $x$'s subtree (excluding the subtree of $x$'s leftmost child that has just been cut) to the pool of trees to be combined. See Figure \ref{fig1}.

\begin{figure}
  \centerline{
\parbox{4in}{\psfig{figure=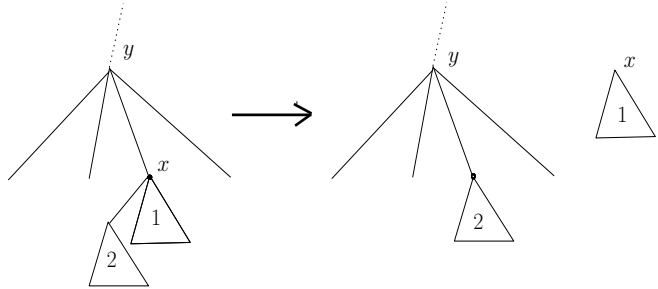,width=3.4in,height=1.5in}}}
\caption{A cut performed by the clean-up procedure.}
\label{fig1}
\end{figure}

\item[ii.] Arbitrary divide the trees of the pool into groups of $\Theta(\log{n})$ trees each (except possibly for one smaller group). For every group, sort the values of the roots of the trees and link the resulting trees in this order such that their roots form a path of nodes in the combined tree (make every root the leftmost child of the root with the next smaller value). Link the combined trees with the main tree in any order. 

\end{enumerate}

\end{description}

\begin{itemize}

\item{\it meld.}
Call clean-up for the smaller heap. Link the trees of the two heaps. Destroy the smaller heap. Update the minimum pointer to point to the root if it has the minimum of the melded heap. 

\item{\it delete-min.}
Call clean-up. Apply the standard two-pass implementation of the pairing heaps \cite{fsst}. 
Make the minimum pointer point to the root of the resulting tree.

\end{itemize}

\section{Analysis}

We prove the following theorem that implies the claimed time bounds:

\begin{theorem}
Starting with an empty heap, consider a sequence of operations $S=\sequence{o_1, o_2,\ldots}$. Let $A = \{i ~|~ o_i$ is a meld operation$\}$, $B = \{i ~|~ o_i$ is a find-min or an insert operation$\}$, $C = \{i ~|~ o_i$ is a decrease-key operation$\}$, and $D = \{i ~|~ o_i$ is a delete-min operation$\}$. The sequence $S$ can be executed on our pairing heaps in $O(|B| + \sum_{i \in C} \log\log{n_i} + \sum_{i \in D} \log{n_i})$, where $n_i$ is the number of elements that are in the heap at operation $i$ and will leave the heap while performing $S$. 
\end{theorem}

For the sake of the analysis, we categorize the nodes as follows. A node is {\it black} if it will remain in the heap after performing the sequence of operations under consideration, otherwise it is {\it white}. A black node whose descendants are all black is called an {\it inactive} node.
Let $w(x)$ be the number of white descendants of a node $x$, including $x$ if it is white.

\begin{enumerate} 
\item {\it Inactive} nodes: Every node $x$ with $w(x) = 0$.
\item {\it Active} nodes: Other nodes. 
\end{enumerate}

To bound the cost of the heap operations, we use a combination of the potential function and the accounting methods \cite{t}.

\subsection{The potential function}  

Consider the link between a node $x$ and its parent $p(x)$.
Let $w'(x)$ be the number of white descendants of $p(x)$ restricted to the subtrees of the right siblings of $x$, including $p(x)$ if it is white. 
We use the potential function
 
\[\Phi = \sum_{w(x) > 0} \log{\frac{w(x)+w'(x)}{w(x)}}.\]

Despite the fact that the potential on a link may reach $\log{n}$, the sum of potentials on a path from a node $z$ to any of its descendants telescopes to at most $\log{w(z)}$. If the path is the left spine of the subtree of $z$, the sum of potentials telescopes to exactly  $\log{w(z)}$.

\subsection{Debits}

Consider the following two cases:
\begin{itemize}
\item a white node is inserted in a heap with an {\it active} root. 
\item two heaps with {\it active} roots are melded. 
\end{itemize}

To fulfill the potential requirements, $O(\sum_{i \in D} \log{n_i})$ units are borrowed from the allowable cost for the {\it delete-min} operations that will be performed on the white nodes. The following lemma illustrates that these debits are enough to cover the above two cases.

\begin{lemma}
\label{l1}
Consider the heap at any time during the execution of the sequence of operations $S$. Let $D' = \{i ~|~ o_i$ is a delete-min operation that will be performed on a node currently in the heap$\}$.
The sum of the potentials on the links formed by insert or meld operations is at most $\sum_{i \in D'} \log{n_i}$, where $n_i$ is the number of elements that are in the heap at operation $i$ and will leave the heap while performing $S$. 
\end{lemma}

{\it Proof.}
Let $\tau$ be a tree representing a heap that has $k>0$ white nodes at this point of time. Let $D'_{\tau}$ be the set $D'$ restricted to the operations performed on the nodes of $\tau$, and $P_{\tau}$ be the sum of the potentials on the links of $\tau$ formed by {\it insert} or {\it meld} operations. We prove by induction the stronger fact that $P_{\tau} \leq \sum_{i=1}^k \log{i}$. Since all the white nodes will eventually be deleted, then $\sum_{i=1}^k \log{i} \leq \sum_{i \in D'_{\tau}} \log{n_i}$.
Consider an {\it insert} operation, where a white node is linked to $\tau$ resulting in the tree $\tau'$. The required potential on this link is $\log{(k+1)}$. By induction, $P_{\tau'} \leq \log{(k+1)} + \sum_{i=1}^k \log{i} =  \sum_{i=1}^{k+1} \log{i}$. 
Consider a {\it meld} operation, where two trees $\tau_1$ and $\tau_2$ with {\it active} roots are linked resulting in tree $\tau'$. Assume that $\tau_1$ and $\tau_2$ have $k_1, k_2 > 0$ white nodes, respectively. The required potential on this link is at most $\log{(k_1+k_2)}$. By induction, $P_{\tau'} \leq \log{(k_1+k_2)} + \sum_{i=1}^{k_1} \log{i} + \sum_{i=1}^{k_2} \log{i} \leq \sum_{i=1}^{k_1+k_2} \log{i}$. This follows from the fact that $k_1! + k_2! < (k_1 + k_2 - 1)!$, for any integers $k_1, k_2 > 1$. 
\hfill $\Box$

\subsection{Credits}

We maintain the following credits in addition to the potential function:

\begin{enumerate}
\item [-] {\it Decrease} credits: $O(\log \log n)$ credits for every decreased node since the previous {\it clean-up} is performed. 
\item [-] {\it Heap} credits: $O(\log{n'})$ credits per heap, where $n'$ is the size of this heap.
\item [-] {\it Active-parent} credits: $O(1)$ credits for every child of an {\it active} node.
\item [-] {\it Active-run} credits: $O(1)$ credits for every {\it active} node with an {\it inactive} right sibling. 
\end{enumerate}

\subsection{The time bounds}

Next, we analyze the time bounds for our operations. Each operation must maintain the potential function, the credits, and pay for the work it performs.

\subsubsection{find-min}
No potential or credit changes are required. The actual work of {\it find-min} is $O(1)$. It follows that the worst-case cost of {\it find-min} is $O(1)$.

\subsubsection{insert} 

If the inserted node is white, extra potential units may be needed. But, as Lemma \ref{l1} illustrates, these units are borrowed from the logarithmic cost per {\it delete-min}, and the {\it insert} operation need not pay for that. 

Assume that as a result of the {\it insert} operation node $x$ is linked to node $y$. If $y$ is {\it active}, the {\it active-parent} credits need to be increased by $O(1)$. If $x$ is {\it active}, and the previous leftmost child of $y$ was {\it inactive}, the {\it active-run} credits need to be increased by $O(1)$. Since the size of the heap increased by one, the {\it heap} credits need to be increased by $O(1)$. The {\it decrease} credits need to be increased by $O(\log \log (n+1) - \log \log{n})$ per decreased node, which still sums up to $O(1)$ as indicated by the following proposition.

\begin{prop}
$n \cdot (\log\log{(n+1)} - \log\log n) = O(1)$.
\end{prop}

{\it Proof.}
For $n>2$,
\begin{eqnarray*}
n \cdot (\log\log{(n+1)} - \log\log n) & < & n \cdot \log{\frac{n + 1}{n}} \\
& = & \log{(1+\frac{1}{n})^n}
\end{eqnarray*}

But $(1+\frac{1}{n})^{n} < e$, where $e$ is the base of the natural logarithm.
\hfill $\Box$ \\

The actual work to link an inserted node with the main tree is $O(1)$. It follows that the amortized cost of {\it insert} is $O(1)$.

\subsubsection{decrease-key}
No potential changes are required. The {\it decrease-key} pays $O(\log \log{n})$ credits for the decreased node. The actual work it performs is $O(1)$. It follows that the amortized cost of the {\it decrease-key} operation is $O(\log\log{n})$.

\subsubsection{clean-up} 

First, consider the effect of a cut performed on a decreased node $x$:

Consider the path of nodes from the root including all the ancestors of $x$ followed by the nodes on the left spine of $x$'s subtree. Since we cut the subtree of $x$ and replace it with the subtree of its leftmost child, the nodes of the above path remain the same except for $x$.
If all the descendants of $x$ are black, possibly excluding the subtree of its leftmost child, then the potentials on all the links do not change as a result of the cut. Otherwise, all the ancestors of $x$ before the cut are {\it active}. 
In such case, the proof given in \cite{elm0}can be applied, indicating that the sum of the potential on all the links does not increase.

If $x$ and both its left and right siblings are {\it active} while its leftmost child is {\it inactive}, then the number of {\it active-runs} increases by one, and $O(1)$ credits would be needed and paid for from the released {\it decrease} credits.

Second, consider the effect of combining the trees and linking them with the main tree: 

The trees of a group are combined by sorting the values in their roots and linking them accordingly in order.  Since the size of a group is $O(\log{n})$, the actual work done in sorting is paid for from the released {\it decrease} credits ($O(\log\log{n})$ credits per node). This will result in a new path of links. Since the sum of the potential values on a path telescopes, the increase in potential as a result of combining the trees of a group and then linking this group to the main tree is $O(\log{n})$. This $O(\log{n})$ potential increase is also paid for from the {\it decrease} credits, except for possibly the last group. (The last group may be a smaller group, and its {\it decrease} credits may not be enough to pay for the increase in potential.) 

As a result of a link the number of {\it active-runs} and {\it active-parents} may increase by one, and $O(1)$ credits would be needed and again paid for from the {\it decrease} credits.  

It follows that the overall amortized cost of the {\it clean-up} procedure is $O(\log{n})$.

\subsubsection{meld}

As for {\it insert}, extra potential units may be needed. But, as Lemma \ref{l1} illustrates, these units are borrowed from the logarithmic cost per {\it delete-min}. 

The cost of the {\it clean-up} performed on the smaller heap is $O(\log{n'})$, where $n'$ is its size. Since the size of the combined heap is at most twice the size of the larger heap, the {\it heap} credits for the combined heap need to be incremented by $O(1)$. Similar to {\it insert}, the {\it active-parent} credits and the {\it active-run} credits may need to be increased by $O(1)$.
The actual work for meld, other than the {\it clean-up} of the smaller heap, is $O(1)$. All these costs are paid for from the {\it heap} credits of the smaller heap, before it is destroyed.     

It follows that the {\it meld} operation pays nothing; everything is taken care of by others.

\subsubsection{delete-min}

\begin{figure}
  \centerline{
\parbox{4in}{\psfig{figure=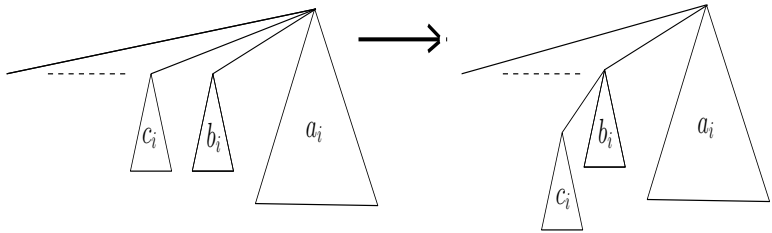,width=4in,height=1.2in}}}
\caption{A delete-min $((c_i \nearrow b_i) \nearrow a_i)$ step.}
\label{fig2}
\end{figure}

We think about the two-pass pairing as being performed in steps. At the $i$-th step, the pair of trees that is the $i$-th pair from the right among the subtrees of the deleted root are linked, then the resulting tree is linked with the combined tree from the linkings of all the previous steps. Each step will then involve three trees and two links. Let $a_i$ be the tree resulting from the linkings of the previous steps, and let $A_i$ be the number of white nodes in $a_i$. Let $b_i$ and $c_i$ be the $i$-th pair from the right among the subtrees of the deleted root to be linked at the $i$-th step, and let $B_i$ and $C_i$ respectively be the number of white nodes in their subtrees. It follows that $A_{i+1}=A_i+B_i+C_i$. Let $(\tau_1 \nearrow \tau_2)$ denote the tree resulting from the linking of tree $\tau_1$ to tree $\tau_2$ as its leftmost subtree. See Figure \ref{fig2}.

We distinguish between four cases, according to the types of the roots of $b_i$ and $c_i$ and who wins the comparison.

\begin{enumerate}

\item Both roots are {\it inactive}:  

There was no potential on the two links that were cut, and no potential is either required on the new links. The actual cost of this step is paid for from the released {\it active-parent} credits, as these two roots were children of an {\it active} parent and at least one of them is not any more. \\

\item An {\it active} root is linked to an {\it inactive} root, and $((\cdots \nearrow \cdots) \nearrow a_i)$: 

The potential that was related to the active root before the operation is enough to cover the potential of the new link with $a_i$.
If the leftmost child of the {\it inactive} root was {\it inactive} before the link, the {\it active-run} credits need to be increased by $O(1)$.  
As for the previous case, these possibly-needed extra credits and the actual cost of the step are paid for from the released {\it active-parent} credits. \\

\item 

\begin{enumerate}

\item Both roots are {\it active}: 

The {\it active-run} credits may need to be increased by $O(1)$.

The potential on the two links that are cut at the $i$-th step was

\[\log{\frac{A_i+B_i}{B_i}} + \log{\frac{A_i+B_i+C_i}{C_i}}.\] 

We consider the four possibilities: \\

\begin{enumerate}

\item $((c_i \nearrow b_i) \nearrow a_i)$:

The potential on the new links is 
\[\log{\frac{B_i+C_i}{C_i}} + \log{\frac{A_i+B_i+C_i}{B_i+C_i}} = \log{\frac{A_i+B_i+C_i}{C_i}}.\]
The difference in potential is 
\[\log{\frac{B_i}{A_i+B_i}} < \log{\frac{B_i+C_i}{A_i}}.\] 
 
\item $((b_i \nearrow c_i) \nearrow a_i)$:

The potential on the new links is
\[\log{\frac{B_i+C_i}{B_i}} + \log{\frac{A_i+B_i+C_i}{B_i+C_i}} = \log{\frac{A_i+B_i+C_i}{B_i}}.\] 
The difference in potential is
\[\log{\frac{C_i}{A_i+B_i}} < \log{\frac{B_i+C_i}{A_i}}.\] 
 
\item $(a_i \nearrow (c_i \nearrow b_i))$:

The potential on the new links is 
\[\log{\frac{B_i+C_i}{C_i}} + \log{\frac{A_i+B_i+C_i}{A_i}}.\] 
The difference in potential is
\[\log{\frac{(B_i+C_i) \cdot B_i}{A_i \cdot (A_i+B_i)}} < 2 \log{\frac{B_i+C_i}{A_i}}.\] 
 
\item $(a_i \nearrow (b_i \nearrow c_i))$:

The potential on the new links is
\[\log{\frac{B_i+C_i}{B_i}} + \log{\frac{A_i+B_i+C_i}{A_i}}.\] 
The difference in potential is
\[\log{\frac{(B_i+C_i) \cdot C_i}{A_i \cdot (A_i+B_i)}} < 2 \log{\frac{B_i+C_i}{A_i}}. \] 

\end{enumerate}

\item One root is {\it active} and the other is {\it inactive}, and $(a_i \nearrow (\cdots \nearrow \cdots))$:

The {\it active-run} credits may need to be increased by $O(1)$. 

Since either $B_i$ or $C_i$ equals zero, we use $M_i = \max{\{B_i, C_i\}}$ for the other value.

The potential on the cut links is
\[\log{\frac{A_i+ M_i}{M_i}}.\]

The potential on the new links is
\[\log{\frac{A_i+M_i}{A_i}}.\] 

The difference in potential is
\[\log{\frac{M_i}{A_i}} = \log{\frac{B_i+C_i}{A_i}}.\] 

\end{enumerate}

\vspace{.1in}

\begin{description}

\item[-] If $B_i+C_i \leq A_i/2$, then $\log{(\frac{B_i+C_i}{A_i})} \leq -1$. Then, for all the above sub-cases, the change in potential is less than $-1$. This released potential is used to pay for the possibly-required increase in the {\it active-run} credits, in addition to the actual work done at this step. 

\item[-] If $B_i+C_i > A_i/2$, we call this step a bad step. For all the above sub-cases, the change in potential resulting from all bad steps is at most $2\sum_i \log{\frac{B_i+C_i}{A_i}}$ (taking the summation for positive terms only, i.e. $B_i+C_i > A_i$). Since $A_{i'} > B_i+C_i$ when $i'>i$, the sum of the changes in potential for all steps telescopes to $O(\log{n})$. It remains to account for the actual work done at the bad steps. Since $A_{i+1}=A_i+B_i+C_i$, a bad step results in $A_{i+1} > \frac{3}{2} A_{i}$.  Then, the number of bad steps is $O(\log{n})$. It follows that the increase in the {\it active-run} credits and the actual work done at bad steps is $O(\log n)$ for each {\it delete-min} operation.\\  

\end{description}

\item An {\it inactive} root is linked to an {\it active} root, and $((\cdots \nearrow \cdots) \nearrow a_i)$: 

The potential that was related to the active root before the operation is enough to cover the potential of the new link with $a_i$. 
To cover the actual work done in such step, consider the two steps that follow it. If those two steps are of the same type as this step, the number of {\it active-runs} decreases (at least one {\it inactive} node is taken out of the way of two {\it active-runs}) and such released credits are used to pay for all three steps (this is similar to Iacono's {\it triple-white} notion in his potential function \cite{ia}). Otherwise, one of those two steps will pay for the current step as well. 

\end{enumerate}

From the above case analysis, it follows that the amortized cost of the {\it delete-min} operation is $O(\log{n})$.

\section{Variations}

The main difference between our implementation and the standard implementation of the pairing heaps is the {\it clean-up} procedure. We chose to perform the {\it clean-up} before the {\it delete-min} operation, and to apply it to the smaller heap before the {\it meld} operation. 
The following variations are as well possible:

\begin{itemize}

\item It is possible to periodically perform the {\it clean-up}, once the number of decreased nodes reaches $\Theta(\log{n})$ following a {\it decrease-key} operation. This assures that when the {\it clean-up} is performed prior to {\it delete-min} operations, there will be only $O(\log{n})$ decreased nodes (one group). 

\item It is possible not to call {\it clean-up} prior to {\it meld} operations, and to do all the work prior to {\it delete-min} operations instead.

\item  In \cite{f}, Fredman stated that the cost of $m$ pairing-heap operations, including $n$ {\it delete-min} operations, is $O(m \log_{2m/n}{n})$. This bound implies a constant cost for the {\it decrease-key} operation when $m=\Omega(n^{1+\epsilon})$, for any constant $\epsilon>0$. This suggests that, when the number of the decreased nodes is large enough, we perform the {\it clean-up} by cutting each of the affected subtrees and directly linking it with the main tree (similar to the standard pairing-heaps implementation). 

\end{itemize}

\section{Conclusion}
We have given a variation of the pairing heaps that achieves the same amortized bounds as Fibonacci heaps, except for {\it decrease-key} (which still matches Fredman's lower bound for, what he calls \cite{f}, a generalized pairing heap). Three important open questions are:

\begin{itemize}
\item Is there a self-adjusting heap that achieves amortized $o(\log \log{n})$ {\it decrease-key} cost?
\item Is it possible that the original implementation of the pairing heaps has the same bounds as those we achieve in this paper?
\item Which heap performs better in practice? 
\end{itemize}

\end{document}